\documentclass[twocolumn,showpacs,preprintnumbers,amsmath,amssymb,superscriptaddress]{revtex4}

\usepackage{graphicx}
\usepackage{dcolumn}
\usepackage{bm}
\usepackage{color}

\begin{document}


\title{Supersonic Air Flow due to Solid-Liquid Impact}

\author{Stephan Gekle}
\author{Ivo R.~Peters}
\affiliation{
Department of Applied Physics and J.M. Burgers
Centre for Fluid Dynamics, University of Twente, P.O. Box 217,
7500 AE Enschede, The Netherlands
}
\author{Jos\'e Manuel Gordillo}
\affiliation{
\'Area de Mec\'anica de Fluidos, Departamento de Ingener\'ia Aeroespacial y Mec\'anica de Fluidos, Universidad de Sevilla, Avda. de los Descubrimientos s/n 41092, Sevilla, Spain
}
\author{Devaraj van der Meer}
\author{Detlef Lohse}
\affiliation{
Department of Applied Physics and J.M. Burgers
Centre for Fluid Dynamics, University of Twente, P.O. Box 217,
7500 AE Enschede, The Netherlands
}

\date{\today}

\begin{abstract}
A solid object impacting on liquid creates a liquid jet due to the collapse of the impact cavity. Using visualization experiments with smoke particles and multiscale simulations we show that in addition a high-speed air-jet is pushed out of the cavity. Despite an impact velocity of only 1~m/s, this air-jet attains \emph{supersonic} speeds already when the cavity is slightly larger than 1~mm in diameter. The structure of the air flow resembles closely that of compressible flow through a nozzle -- with the key difference that here the ``nozzle'' is a \emph{liquid} cavity shrinking rapidly in time.
\end{abstract}


\maketitle


Taking a stone and throwing it onto the quiescent surface of a lake triggers a spectacular series of events which has been the subject of scientists' interest for more than a century \cite{Worthington_book_1908, GilbargAnderson_JApplPhys_1948, May_JApplPhys_1951, May_JApplPhys_1952, Abelson_JFM_1970, GlasheenMcMahon_PhysFluids_1996, LeeLongoriaWilson_PhysFluids_1997, Gaudet_PhysFluids_1998, BergmannEtAl_PRL_2006, DuclauxEtAl_JFM_2007, VellaMetcalfe_PhysFluids_2007, DuezEtAl_NaturePhys_2007, GekleEtAl_PRL_2008, GekleEtAl_PRL_2009, AristoffBush_JFM_2009, BergmannEtAl_JFM_2009, DoQuangAmberg_PhysFluids_2009}: upon impact a thin sheet of liquid (the ``crown splash'') is thrown upwards along the rim of the impacting object while below the water surface a large cavity forms in the wake of the impactor. Due to the hydrostatic pressure of the surrounding liquid this cavity immediately starts to collapse and eventually closes in a single point ejecting a thin, almost needle-like liquid jet. Just prior to the ejection of the liquid jet the cavity possesses a characteristic elongated ``hourglass'' shape with a large radius at its bottom, a thin neck region in the center, and a widening exit towards the atmosphere.


This shape is very reminiscent of the converging-diverging (``de Laval'') nozzles known from aerodynamics as the paradigmatic picture of compressible gas flow through, e.g.,  supersonic jet engines. In this Letter we use a combination of experiments and numerical simulations to show that in addition to the very similar shape, also the structure of the air flow through the impact cavity resembles closely the high-speed flow of gas through such a nozzle. Not only is the flow to a good approximation one-dimensional, but it even attains supersonic velocities. Nevertheless, the pressure inside the cavity is merely 2\% higher than the surrounding atmosphere.
The key difference, however, is that in our case the ``nozzle'' is a liquid cavity whose shape is evolving rapidly in time  -- a situation for which no equivalent exists in the scientific or engineering literature.

Our experimental setup consists of a thin circular disc with radius $R_0=2$ cm which is pulled through the liquid surface by a linear motor mounted at the bottom of a large water tank \cite{BergmannEtAl_JFM_2009} with a constant speed of $V_0=1$~m/s. To visualize the air flow we use small glycerin droplets (diameter roughly 3~$\mu$m) produced by a commercially available smoke machine (skytec) commonly used for light effects in theaters and discotheques. Before the start of the experiment the atmosphere above the water surface is filled with this smoke which is consequently entrained into the cavity by the impacting disc. A laser sheet (Larisis Magnum II, 1500mW) shining in from above illuminates a vertical plane containing the axis of symmetry of the system. A high-speed camera (Photron SA1.1) records the motion of the smoke particles at up to 15,000 frames per second. Cross-correlation of subsequent images allows us to extract the velocity of the smoke which faithfully reflects the actual air speed \cite{EPAPS_AirFlow_1}. Our setup obeys axisymmetry and we use cylindrical coordinates with $z=0$ the level of the undisturbed free surface.

\begin{figure}
\includegraphics[width=0.32\columnwidth]{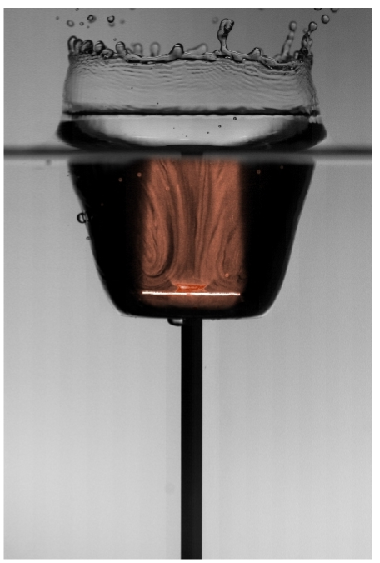}
\includegraphics[width=0.32\columnwidth]{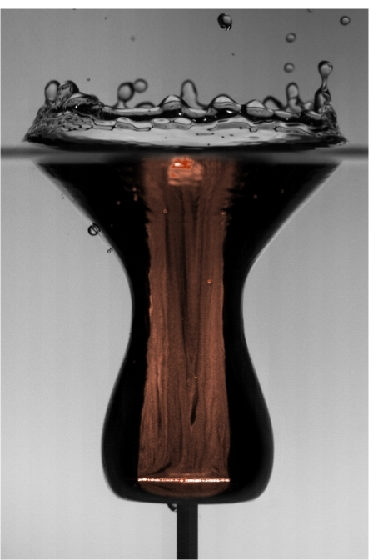}
\includegraphics[width=0.32\columnwidth]{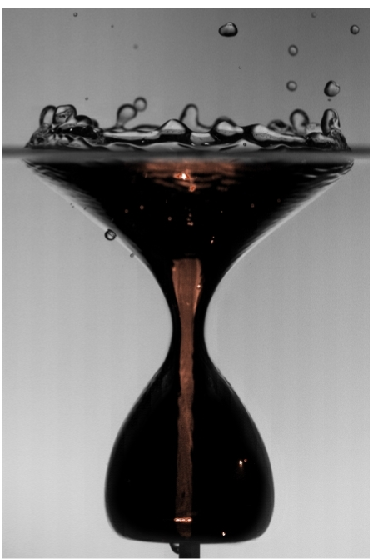}
\caption{(a) After the impact of the disc an axisymmetric cavity is formed in its wake and air is entrained into this cavity. (b) Due to hydrostatic pressure from the surrounding liquid the cavity starts to collapse and the air flow reverses its direction. (c) As the collapse proceeds air is pushed out of the shrinking cavity at very high speeds. In (a)--(c) we overlaid images of the cavity shape (recorded with backlight) and images of the smoke particles (recorded with the laser sheet and artificially colored in orange). In the latter, the area illuminated by the vertical laser sheet is restricted by the minimum cavity radius \cite{EPAPS_AirFlow_1}. A corresponding movie can be found in \cite{EPAPS_AirFlow_2}.}
\label{fig:sequence_airFlow}
\end{figure}
\begin{figure}
\begin{center}
\includegraphics[width=\columnwidth]{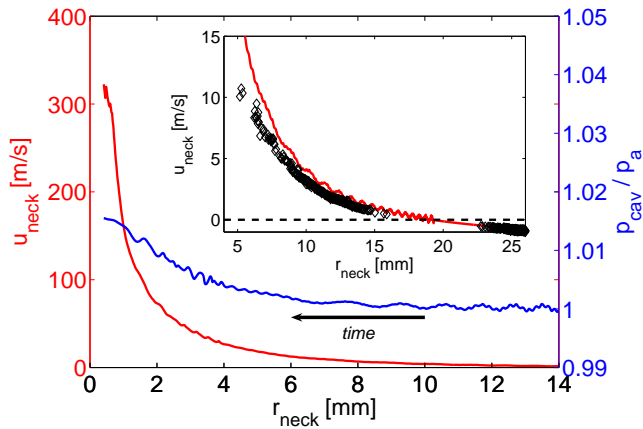}
\caption{The speed of the gas flowing through the neck (red curve) as a function of the shrinking cavity neck taken from the fully compressible simulations. The main plot demonstrates that sonic speeds are attained with the cavity pressure (blue curve) being less than 2\% higher than the atmospheric pressure. The enlargement (inset) shows that the numerical scheme (red curve) agrees very well with the experimentally measured velocity (black diamonds; the hole in the data between $r_\mathrm{neck}=16$~mm and 22 mm is due to measurement uncertainties at low absolute velocities \cite{EPAPS_AirFlow_1}). Slight non-axisymmetric perturbations \cite{SchmidtEtAl_NaturePhys_2009, TuritsynLaiZhang_PRL_2009} in the experimental setup may be responsible for the somewhat slower air speed of the experiment as compared to the simulation. One can clearly see the inversion of the flow direction from negative (into the cavity) to positive (out of the cavity) velocities.}
\label{fig:velocityGas}
\end{center}
\end{figure}

In the beginning of the process (see the snapshot in Fig.~\ref{fig:sequence_airFlow}~(a)) air is drawn into the expanding cavity behind the impacting object with velocities of the order of the impact speed. In a later stage however, this downward flux is overcompensated by the overall shrinking of the cavity volume resulting in a net flux out of the cavity. The cavity shape at the moment when the flow through the neck reverses its direction is illustrated in Fig.~\ref{fig:sequence_airFlow}~(b). Towards the end of the cavity collapse a thin and fast air stream is pushed out through the cavity neck which is illustrated in Fig.~\ref{fig:sequence_airFlow}~(c). From images such as those in Fig.~\ref{fig:sequence_airFlow} we can directly measure the air speed $u$ up to about 10~m/s as is shown in the inset of Fig.~\ref{fig:velocityGas}.

In order to determine the flow speed at even higher velocities we revert to multiscale numerical simulations. Our numerical method proceeds in two stages: an incompressible stage at the beginning and a compressible stage towards the end of the impact process. During the first stage both air and liquid are treated as incompressible, irrotational, and inviscid potential fluids. To solve for the flow field and to calculate the motion of the interface we use a boundary integral method (BIM) as described in \cite{BergmannEtAl_JFM_2009} with extensions to include the gas phase \cite{RodriguezRodriguezGordilloMartinezBazan_JFM_2006}.
At the moment that the air flow through the neck reverses, see Fig.~\ref{fig:sequence_airFlow}~(b), the simulation enters into the second, compressible stage: from now on only the liquid motion is computed by the incompressible BIM.

To simulate the air flow in the second stage we need to take compressibility into account meaning that a simple potential flow description is no longer possible. Fortunately, at the end of the incompressible stage the air velocity profile is almost perfectly one-dimensional along the axis of symmetry. We can therefore describe the gas dynamics by the 1D compressible Euler equations \cite{Laney_book_1998} in analogy to gas flowing through a converging-diverging nozzle. In the Euler equations we include two additional terms accounting for the variation of the nozzle radius in time and space \cite{EPAPS_AirFlow_1}. For the numerical solution we use a Roe scheme \cite{Roe_JComputPhys_1981, Laney_book_1998} which is highly appreciated for its computational efficiency and ability to accurately capture shock fronts.

The two-way coupling between the gas and the liquid domains is accomplished via (i) the interfacial shape and its instantaneous velocity which is provided by the BIM and serves as an input into the gas solver and (ii) the pressure which is obtained from the solution of the Euler equations and serves as a boundary condition for the BIM. Above the location of the initial free surface the surface pressure of the BIM remains atmospheric.

Combining our experiments with these numerical simulations leads to the main result of this Letter contained in Fig.~\ref{fig:velocityGas}: the collapsing liquid cavity acts as a rapidly deforming nozzle, so violent that the air which is pushed out through the neck attains supersonic velocities (red line). Our simulations show that the pressure inside the cavity which is driving this flow is less than 1.02 atmospheres (blue line). From the inset one can tell that our simulations are in good agreement with the smoke measurements over the entire experimentally accessible range. It is interesting to note that even towards the end of the process (when sonic velocities are reached) there is a net flux of air upwards through the cavity. If the process was governed merely by the collapse of the neck itself one would expect the air to be pushed out of the neck region in both vertical directions. This net flow thus underlines the important role of the dynamics of the entire cavity.

\begin{figure}
\begin{center}
\includegraphics[width=\columnwidth]{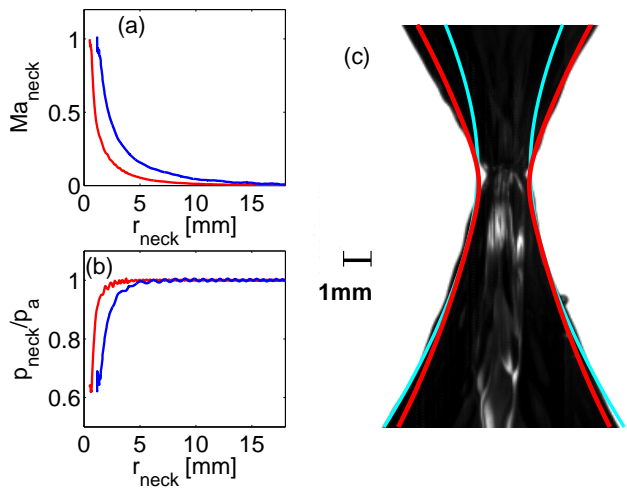}
\caption{(a) The evolution of the local Mach number at the cavity neck for different impact speeds (red: 1 m/s, blue: 2 m/s). For the 2 m/s impact speed sonic flow is attained at a cavity radius of 1.2 mm. (b) The pressure at the neck diminishes due to Bernoulli suction as the neck radius shrinks and air is forced to flow faster and faster. The minimum pressure lies at about $0.6 p_a$ which is attained when the Mach number reaches unity. (c) The experimental image shows a pronounced kink at the neck which is not captured by the smoothly rounded curve predicted by the simulation without air (cyan line). Only the inclusion of air effects into the simulations (red line) is able to reproduce the kinked shape caused by the low air pressure at the neck as well as the shape of the cavity above the neck.}
\label{fig:pressureNeck}
\end{center}
\end{figure}

To determine more precisely at what point the air flow through the neck becomes sonic we show in Fig.~\ref{fig:pressureNeck}~(a) the evolution of the local Mach number, $\mathrm{Ma_{neck}} = u_\mathrm{neck}/c$ (with the gas velocity $u_\mathrm{neck}$ and the speed of sound $c$), for discs impacting at 1 and 2 m/s. We find that the speed of sound is attained at cavity radii as large as 0.5 mm for the lower impact velocity and 1.2~mm for the higher impact velocity.

In a steady state one could expect from the (compressible) Bernoulli equation that these very high air speeds would cause a greatly diminished air pressure in the neck region. Despite the unsteadiness of our situation, the data presented in Fig.~\ref{fig:pressureNeck}~(b) indeed shows that the pressure $p_\mathrm{neck}$ decreases significantly once the neck has shrunk to a diameter of roughly 4 mm (for the 1 m/s impact) while before that point it is practically atmospheric throughout. Classical steady-state theory \cite{LiepmanRoshko_book_1957} for a converging-diverging nozzle predicts that when $\mathrm{Ma_{neck}} = 1$ the pressure at the neck reaches a minimum value of
\begin{equation}
p_\mathrm{neck} / p_a = \left(1 + \frac{\gamma-1}{2} \right)^{-\gamma/(\gamma-1)}  = 0.53
\end{equation}
with $p_a$ the atmospheric pressure and $\gamma = 1.4$ the isentropic exponent. As shown in Fig.~\ref{fig:pressureNeck}~(b) our situation -- although highly unsteady -- exhibits a similar behavior with $p_\mathrm{neck}\approx 0.6p_a$ as the Mach number becomes of order unity.

In Fig.~\ref{fig:pressureNeck}~(c) we illustrate how this low pressure gives us a handle to observe the consequences of the supersonic air speed in our experiments: despite the air being three orders of magnitude less dense than water, it is able to exert a significant influence even on the shape of the liquid cavity provided that its speed is high enough \cite{GordilloEtAl_PRL_2005, BergmannEtAl_PRL_2009}. From the experimental image it is clear that the free surface close to collapse no longer possesses a smoothly rounded shape but instead shows a significant increase in curvature at the minimum (a ``kink''). While this feature is not present in a simulation neglecting the influence of air as those in \cite{BergmannEtAl_JFM_2009}, the inclusion of air effects allows us to capture quite accurately the cavity shape observed experimentally. This gives strong evidence that in the experiment the air indeed becomes as fast as predicted by the simulations and produces a Bernoulli suction effect strong enough to deform the cavity.

\begin{figure}
\begin{center}
\includegraphics[width=\columnwidth]{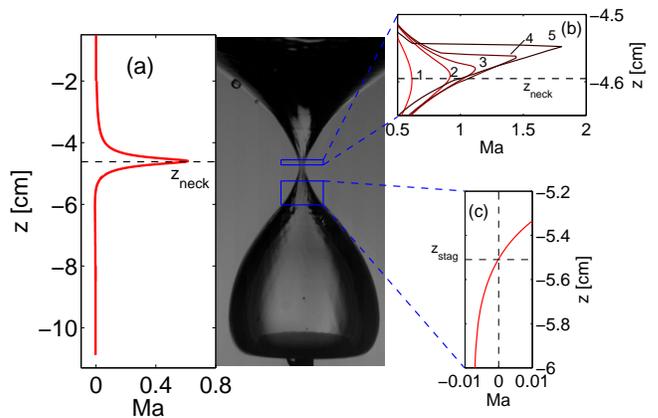}
\caption{(a) The vertical air velocity normalized by the local speed of sound  $\mathrm{Ma} = u / c$ as a function of the vertical position (the corresponding cavity image is shown in the middle) for $r_\mathrm{neck} = 0.9$~mm: the profile exhibits a sharp peak approximately at the height of the neck. (b) A close-up of the zone around the neck illustrates the steepening of the velocity profiles towards pinch-off (numbers 1-5 correspond to neck radii between 0.9 mm (number 1, bright red) and 0.5 mm (number 5, dark brown)) and the development of the shock front at roughly 0.1 ms before pinch-off. The neck position $z_\mathrm{neck}$ corresponding to curve 5 is shown by the dashed line. (c) A close-up of the area below the neck shows the location of the gas flow stagnation point $z_\mathrm{stag}$ (dashed line).}
\label{fig:airVelocityProfile}
\end{center}
\end{figure}

The positive sign of $u_\mathrm{neck}$ (see Fig.~\ref{fig:velocityGas}) indicates that the gas flow is directed upwards at the neck. At the same time, the air at the bottom of the cavity is pulled downwards by the moving disc. An interesting consequence of this competition between cavity expansion at the bottom and cavity shrinking in the neck is the existence of a stagnation point with $u = 0$ as can readily be observed in Fig.~\ref{fig:airVelocityProfile}~(a) and its magnification in Fig.~\ref{fig:airVelocityProfile}~(c).

\begin{figure}
\begin{center}
\includegraphics[width=\columnwidth]{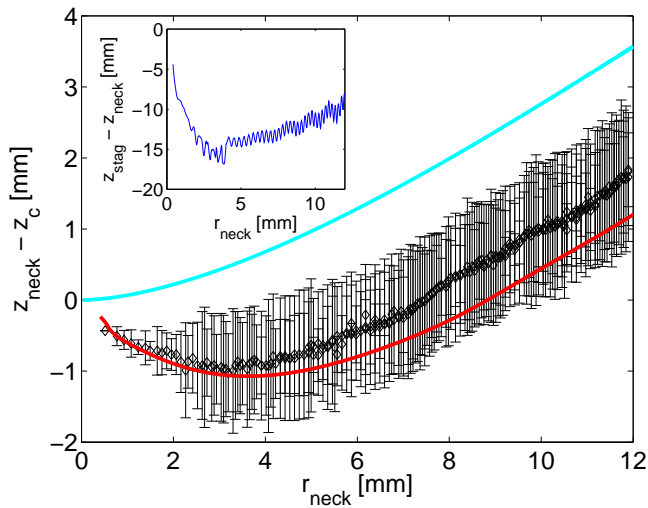}
\caption{The vertical position of the cavity neck relative to the final closure height $z_c$ as a function of the shrinking neck radius from experiment (black diamonds), simulations with (red line) and without (cyan line) air dynamics. The experimental data is in quantitative agreement with the compressible simulations, while clearly the simulation neglecting air fails to capture the upward motion of the minimum induced by the large pressure gradient between the stagnation point and the cavity neck. Experimental error bars are determined by the number of vertically neighboring pixels all sharing the same minimum radius. The inset shows the approach of the stagnation point to the neck.}
\label{fig:neckZPos}
\end{center}
\end{figure}

As can be seen in the inset of Fig.~\ref{fig:neckZPos}, the distance between the neck and the stagnation point is no larger than roughly 5 mm prior to cavity closure. Nevertheless, the pressure at the stagnation point equals the overall pressure inside the cavity which is roughly atmospheric during the whole process (see Fig.~\ref{fig:velocityGas}). Recalling that $p_\mathrm{neck} \approx 0.6p_a$ this results in a tremendous vertical pressure gradient which of course affects the dynamics of the cavity wall: the flow of air is so strong that it can drag the liquid along resulting in an upward motion of the cavity neck just before the final collapse. That this effect is indeed present in the simulations can be seen from the red line in Fig.~\ref{fig:neckZPos}. For comparison, the cyan curve demonstrates that a single fluid simulation neglecting the air dynamics would predict a monotonously decreasing position. The experimental data however is in quantitative agreement with the compressible simulations. Together with the cavity shape shown in Fig.~\ref{fig:pressureNeck}~(c) these results constitute an impressive -- albeit indirect -- demonstration of the credibility of our numerical predictions despite the fact that, understandably, it is not possible to directly measure (super-)sonic air speeds with our smoke setup. Furthermore they show that the perfectly axisymmetric approach of the simulations is justified and, therefore, that supersonic gas velocities are reached before instabilities \cite{SchmidtEtAl_NaturePhys_2009, TuritsynLaiZhang_PRL_2009} inevitably destroy the axisymmetry of the system.

Looking more closely at the velocity profile above the neck (see Fig.~\ref{fig:airVelocityProfile}~(b)) one finds that it possesses a discontinuous jump: the signature of a shock front developing in the air stream. While such a shock front is a common phenomenon in steady supersonic flows, here we are able to illustrate its development even in our highly unsteady situation when the gas velocity passes from sub- to supersonic.

In conclusion, we showed that the air flow inside the impact cavity formed by a solid object hitting a liquid surface attains supersonic velocities. We found that the very high air speeds can be reached even though the pressure inside the cavity is merely 2\% higher than the surrounding atmosphere. This is due to the highly unsteady gas flow created by the rapidly deforming cavity. We illustrated how the air affects the cavity shape close to the final collapse in two different ways: (i) the initially smoothly curved neck shape acquires a kink which can be attributed to a Bernoulli suction effect and (ii) the initially downward motion of the neck reverses its direction and starts to travel upwards. The quantitatively consistent observation of both effects in numerics and experiment makes us confident that our rather involved numerical procedure truthfully reflects reality.

We thank A.~Prosperetti , J.~Snoeijer, and L.~van~Wijngaarden  for discussions. This work is part of the program of the Stichting FOM, which is financially supported by NWO. JMG thanks the financial support of the Spanish Ministry of Education under project DPI2008-06624-C03-01.

\end{document}